\documentclass[universe,article,accept,pdftex,moreauthors]{Definitions/mdpi}
\firstpage{1}
\pubvolume{11}
\issuenum{06}
\articlenumber{171}
\pubyear{2025}
\copyrightyear{2025}
\externaleditor{Stefano Vercellone} 
\datereceived{1 April 2025}
\daterevised{18 May 2025} 
\dateaccepted{20 May 2025}
\datepublished{27 May 2025}
\hreflink{https://doi.org/10.3390/\linebreak  universe11060171} 

\newcolumntype{C}[1]{>{\PreserveBackslash\centering}m{#1}}
\newcolumntype{R}[1]{>{\PreserveBackslash\raggedleft}m{#1}}
\newcolumntype{L}[1]{>{\PreserveBackslash\raggedright}m{#1}}

\Title{Optical Photometric Monitoring of the Blazar OT 355 and Local Standard Stars' Calibration}
\TitleCitation{Optical Photometric Monitoring of the Blazar OT 355 and Local Standard Stars' Calibration}

\Author{{R. Bachev} 
$^{1,}${*}, Tushar Tripathi $^{2}$\href{https://orcid.org/0009-0006-3586-2489}{\orcidicon}, Alok C. Gupta $^{2}$\href{https://orcid.org/0000-0002-9331-4388}{\orcidicon}, A. Kurtenkov $^{1,3}$, Y. Nikolov $^{1}$\href{https://orcid.org/0000-0003-0317-3866}{\orcidicon}, A. Strigachev $^{1}$, S. Boeva $^{1}$,\linebreak   G. Latev $^{1}$\orcidD{}, B. Spassov $^{1}$\href{https://orcid.org/0009-0005-8249-8619}{\orcidicon}, M. Minev $^{1}$\href{https://orcid.org/0000-0001-5394-8258}{\orcidicon}, E. Ovcharov $^{4}$, W.-X. Yang $^{5,6,7,8}$, Yi Liu $^{5,7,8}$\orcidF{} and J.-H. Fan $^{5,7,8}$ \href{https://orcid.org/0000-0003-3863-9777}{\orcidicon}}
\AuthorNames{R. Bachev, Tushar Tripathi, Alok C. Gupta, A. Kurtenkov, Y. Nikolov, A. Strigachev, S. Boeva, G. Latev,  B. Spassov, M. Minev, E. Ovcharov, W.-X. Yang, Yi Liu  and J.-H. Fan}

\AuthorCitation{{Bachev,} 
R.; Tripathi, T.; Gupta, A.C.; Kurtenkov, A.; Nikolov, Y.; Strigachev, A.; Boeva, S.; Latev, G.; Spassov, B.; Minev, M.; et al.}

\address{$^1$ \quad Institute of Astronomy and NAO, Bulgarian Academy of Sciences, 72 Tsarigradsko Shose, 1784 Sofia, Bulgaria;  {al.kurtenkov@gmail.com (A.K.); ynikolov@astro.bas.bg (Y.N.); anton@nao-rozhen.org (A.S.); \linebreak  sboeva@astro.bas.bg (S.B.); glatev@astro.bas.bg (G.L.); b2\_bgb2@yahoo.com (B.S.);\linebreak   mminev@astro.bas.bg (M.M.)} 
\\
$^2$ \quad Aryabhatta Research Institute of Observational Sciences (ARIES), Manora Peak, Nainital 263001, India; {tushar22594@gmail.com (T.T.);  acgupta30@gmail.com (A.C.G.)} \\
$^3$ \quad Faculty of Physics, Sofia University ``St. Kliment Ohridski'', 5 James Bourchier Blvd., 1164 Sofia, Bulgaria \\
$^4$ \quad Department of Astronomy, Faculty of Physics, Sofia University ``St. Kliment Ohridski'', {1116} {Sofia,} 
Bulgaria; {evgeni@phys.uni-sofia.bg}\\
$^5$ \quad Center for Astrophysics, Guangzhou University, Guangzhou 510006,  China; {wenxin.yang@unipd.it (W.-X.Y.); pinux@gzhu.edu.cn (Y.L.); fjh@gzhu.edu.cn (J.-H.F.)}\\
$^6$ \quad Dipartimento di Fisica e Astronomia ``G. Galilei'', Universit'a di Padova,  35131 Padova, Italy\\
$^7$ \quad Astronomy Science and Technology Research {Laboratory} 
of Department of Education of Guangdong Province, Guangzhou 510006,  China\\
$^8$ \quad Greater Bay Brand Center of the National Astronomical Data {Center}, Guangzhou 510006,  China}

\corres{Correspondence: bachevr@astro.bas.bg}

\abstract{OT 355 ({4}FGL J1734.3 + 3858) is a relatively rarely studied but highly variable, {moderate}-redshift (z = 0.975) flat-spectrum radio quasar (blazar). With this work, we aim to study its optical variability on different timescales, which can help us to better understand the physical processes in relativistic jets operating in blazar-type active galactic nuclei. OT 355 was observed in four colors (BVRI) {during 41} nights between 2017 and 2023 using three 1 and 2 m class telescopes. The object was also monitored on intra-night timescales, for about 100 h in total. In addition, secondary standard stars in the field of OT 355 were calibrated in order to facilitate future photometric studies. We detected significant intra-night and night-to-night variations of up to 0.5 mag. Variability characteristics, color changes, and a possible ``rms-flux'' relation were studied and discussed.  Using simple arguments, we show that a negative ``rms-flux'' relation should be expected if many independent processes/regions drive the short-term variability via Doppler factor changes, which is not observed in this and other cases. {This finding raises arguments for the idea that more complex multiplicative processes are responsible for blazar variability.}
Studying blazar variability, especially on the shortest possible timescales, can help to estimate the strength and geometry of their magnetic fields, the linear sizes of the emitting regions, and other aspects, which may be of importance for constraining and modeling blazars' \mbox{emitting mechanisms.}}

\keyword{blazars; variability; relativistic jets}

\begin{document}



\section{Introduction}

{Blazars are active galactic nuclei (AGNs) with almost entirely non-thermal emissions that are produced via synchrotron and inverse Compton (IC) processes in a relativistic jet.  Blazars are classified as jet-dominated AGNs whose jets are pointed at a close angle with respect to the observer, such that relativistic effects can significantly boost the observed radiation (both spectrally and energetically). Their spectral energy distribution (SED) covers almost the entire accessible electromagnetic (EM) spectrum and consists of two humps, which are attributed to synchrotron (the low-energy hump) and IC processes (the high-energy hump). There is a well-known anti-correlation between the peak luminosity and the peak frequency of the humps, which is known as the ``blazar sequence'' \citep{Fos98, Pra22} {and} 
is yet to be properly explained. A useful classification divides blazars into three groups---low-spectrum peaked (LSP), intermediate-spectrum peaked (ISP), and high-spectrum peaked (HSP)---depending on the exact synchrotron peak frequency ($\nu_{\rm syn}^{\rm peak}$) range within which they fall. Thus,  if $\nu_{\rm syn}^{\rm peak} \leq 10^{14}$ Hz, the blazars are considered LSP; \mbox{if $10^{14} \leq \nu_{\rm syn}^{\rm peak} \leq 10^{15}$ Hz,} then the blazars are ISP; and if $\nu_{\rm syn}^{\rm peak} \geq 10^{15}$ Hz, then they are HSP  \citep{Abdo10}.}

Unlike other AGNs, blazars do not show prominent emission lines except for a sub-class of objects that are typically associated with the LSP group; these are dubbed flat-spectrum radio quasars (FSRQs). FSRQs possess the characteristics of both a quasar (thin accretion disk and line-emitting regions) and a blazar (relativistic jet). They are also known to be statistically more optically active than other blazars, showing more prominent and more rapid variations (e.g., \citep{Hov14, Zha15, Bach18, Anj20, Zha24}) on both long and short (even \mbox{intra-night) timescales. }

Blazars are well known for their huge and rapid variability in all energy bands; some of the most prominent objects still retain their variable stars' names (e.g., BL Lacertae, AP Librae, etc.). Different scenarios have been proposed throughout the years to explain the variability of blazars, but an entirely consistent picture has yet to emerge. Understandably, the physics of  relativistic jets can be very complicated; processes such as standing or propagating shocks and ``plasmoids'' moving along relatively ordered (helical) or highly disordered (turbulent) magnetic fields within a jet are to be considered (e.g., \citep{Bot13, Mar14, Mar15, Zha14}). Additionally, a changing Doppler factor (defined in Section \ref{sec4}) of the emitting region that is due to a motion along a curve (i.e., a curved magnetic field), jet precession, and so on can produce a seemingly variable flux at the observer's position, even though the intrinsic energy production may remain relatively stable (e.g., \citep{Rai23}).

In this work, we study the optical variability of OT~355 (3FGL J1734.3 + 3858)---a {moderate}-redshift FSRQ at $z=0.975$ with a central black hole mass of about 9 $\times$ 10$^{7} \rm{M}_{\odot}$~\citep{2012ApJ...748...49S}. The object was observed in four colors (BVRI) on more than 35 nights between  2017 and 2023 with the telescopes of Belogradchik Observatory and Rozhen NAO, Bulgaria. The object was monitored on intra-night (or intra-day, IDV) timescales for about 100 h in total. We were mostly interested in building a statistically significant ``\textit{rms}-flux'' relation for the intra-night variability, identifying the shortest variability timescales, studying the color changes, and building the long-term structure function (SF) of this object's variability, among other things. The next sections provide details of our observations and results, as well as a short discussion.

\section{Observations}
The observations of the flat-spectrum radio quasar (FSRQ) OT 355 started in June 2017, as the object experienced an outburst in the optical {band} 
\citep{Bach17a}. During this outburst, the object reached about the 15th magnitude, while it typically showed between the 16th and 21st magnitude (CRTS {survey} 
\endnote{\url{http://crts.caltech.edu/}  {(accessed on 26 May 2025)}.}). 
As OT~355 is a rarely studied object, we decided to monitor it more thoroughly, which included observing it in different colors and on intra-night timescales. Our data were acquired with the \mbox{60 cm} telescope of Belogradchik Observatory {\citep{Str11}}, as well as the 200 cm and the \mbox{50/70 cm} telescopes of Rozhen National {Observatory} \endnote{\url{http://nao-rozhen.org/}  (accessed on 26 May 2025).}, Bulgaria, all of which were equipped with CCDs and standard UBVRI filter sets. As the object was fairly weak, we performed most of our intra-night observations using no filter in order to improve the signal-to-noise ratio {by almost two times}. In addition, standard stars were calibrated by observing secondary standards in nearby fields.

A log of observations is presented in Table \ref{tab:log}, where the JD of the observation, the evening date, the duration of the observation (if it is a time series, not a single estimate), the filter used for that time series (E = empty; i.e., unfiltered light), the average R-magnitude, and the fractional variability (or \textit{rms}; see the next section) of the intra-night light curve (where applicable) are presented.

\begin{table}[H]
\small
\caption{Log of observations.}
\label{tab:log}
\begin{tabularx}{\textwidth}{C{2.1cm}C{1.9cm}C{1.8cm}C{1.2cm}C{1.2cm}C{1.2cm}C{1.5cm}}
\toprule
\textbf{JD}	&	\textbf{Evening}	&	\textbf{Duration [h]}	&	\textbf{Filter}	&	\boldmath{$<R>$}	&	\boldmath{$rms$}	&	\textbf{Telescope}	\\
\midrule
2,457,924.368	&	19.06.2017
&	2.1	&	E	&	15.75	&	0.04	&	B60	\\
2,457,925.349	&	20.06.2017	&	3.8	&	E	&	16.17	&	0.05	&	B60	\\
2,457,926.424	&	21.06.2017	&	2.3	&	R	&	16.33	&	0.02	&	R200	\\
2,457,927.347	&	22.06.2017	&	3.3	&	E	&	16.11	&	0.02	&	B60	\\
2,457,929.351	&	24.06.2017	&	2.7	&	E	&	15.62	&	0.04	&	B60	\\
2,457,929.492	&	24.06.2017	&	3.0	&	R	&	15.68	&	0.03	&	R200	\\
2,457,930.448	&	25.06.2017	&	2.3	&	R	&	15.17	&	0.04	&	R50/70	\\
2,457,932.365	&	27.06.2017	&	1.8	&	V	&		&	0.02	&	R200	\\
2,457,932.360	&	27.06.2017	&	1.7	&	I	&		&	0.04	&	B60	\\
2,457,932.355	&	27.06.2017	&	1.7	&	R	&	16.30	&	0.02	&	R50/70	\\
2,457,934.399	&	29.06.2017	&	2.8	&	V	&		&	0.01	&	R200	\\
2,457,935.334	&	30.06.2017	&	2.0	&	R	&	16.10	&	0.03	&	R50/70	\\
2,457,936.374	&	01.07.2017	&	2.5	&	R	&	16.56	&	0.01	&	R50/70	\\
2,457,955.319	&	20.07.2017	&	4.5	&	E	&	16.39	&	0.05	&	B60	\\
2,457,956.308	&	21.07.2017	&	3.0	&	E	&	15.77	&	0.05	&	B60	\\
2,457,958.357	&	23.07.2017	&	5.5	&	E	&	16.56	&	0.07	&	B60	\\
2,457,959.337	&	24.07.2017	&	1.0	&	E	&	16.75	&	0	&	B60	\\
2,457,960.353	&	25.07.2017	&	1.0	&	E	&	16.53	&	0.03	&	B60	\\
2,457,964.331	&	29.07.2017	&	$-$	&	$-$	&	15.89	&	$-$	&	R50/70	\\
2,457,965.452	&	30.07.2017	&	$-$	&	$-$	&	16.62	&	$-$	&	R50/70	\\
2,457,966.321	&	31.07.2017	&	$-$	&	$-$	&	16.72	&	$-$	&	R50/70	\\
2,457,975.289	&	09.08.2017	&	5.5	&	E	&	16.44	&	0.03	&	B60	\\
2,457,976.293	&	10.08.2017	&	4.0	&	E	&	16.27	&	0.01	&	B60	\\
2,457,977.284	&	11.08.2017	&	6.0	&	E	&	16.83	&	0.02	&	B60	\\
2,457,994.325	&	28.08.2017	&	$-$	&	$-$	&	17.02	&	$-$	&	B60	\\
2,457,995.359	&	29.08.2017	&	$-$	&	$-$	&	16.44	&	$-$	&	B60	\\
2,458,012.254	&	15.09.2017	&	$-$	&	$-$	&	16.96	&	$-$	&	B60	\\
2,458,247.473	&	08.05.2018	&	2.2	&	R	&	16.35	&	0.03	&	B60	\\
2,458,317.403	&	17.07.2018	&	3.5	&	E	&	16.38	&	0.02	&	B60	\\
2,458,319.402	&	19.07.2018	&	5.0	&	E	&	16.38	&	0	&	B60	\\
2,458,320.365	&	20.07.2018	&	2.1	&	E	&	16.72	&	0.03	&	B60	\\
2,458,341.344	&	10.08.2018	&	2.2	&	E	&	16.91	&	0	&	B60	\\
2,458,342.362	&	11.08.2018	&	2.0	&	E	&	16.83	&	0.02	&	B60	\\
2,458,673.389	&	08.07.2019	&	2.0	&	R	&	15.90	&	0	&	B60	\\
2,458,720.294	&	24.08.2019	&	2.1	&	E	&	15.65	&	0.01	&	B60	\\
2,458,721.282	&	25.08.2019	&	3.2	&	E	&	15.76	&	0.03	&	B60	\\
2,458,724.270	&	28.08.2019	&	3.1	&	E	&	15.81	&	0.01&	B60	\\
2,458,964.565	&	24.04.2020	&	$-$	&	$-$	&	19.34	&	$-$	&	B60	\\
2,458,992.362	&	22.05.2020	&	$-$	&	$-$	&	19.14	&	$-$	&	B60	\\
2,459,029.330	&	28.06.2020	&	$-$	&	$-$	&	18.68	&	$-$	&	R200	\\
2,459,029.355	&	28.06.2020	&	3.0	&	R	&	18.77	&	0	&	R200	\\
2,459,059.509	&	28.07.2020	&	$-$	&	$-$	&	18.97	&	$-$	&	R200	\\
2,460,137.322	&	11.07.2023	&	$-$	&	$-$	&	17.27	&	$-$	&	B60	\\
2,460,195.267	&	07.09.2023	&	$-$	&	$-$	&	18.57	&	$-$	&	B60	\\

\bottomrule
\end{tabularx}
\noindent{\footnotesize{{Telescope codes:} 
} B60: 60-cm Cassegrain telescope, Belogradchik Astronomical Observatory, Bulgaria. R200: \mbox{200-cm} RCC telescope, Rozhen National Astronomical Observatory, Bulgaria. R50/70: 50/70-cm Schmidt telescope, Rozhen National Astronomical Observatory, Bulgaria.}

\end{table}

\section{Methods of Analysis and Results}

\subsection{Secondary Standards}
\begin{figure}[H]
\label{fig:ot355}
\includegraphics[width=85mm]{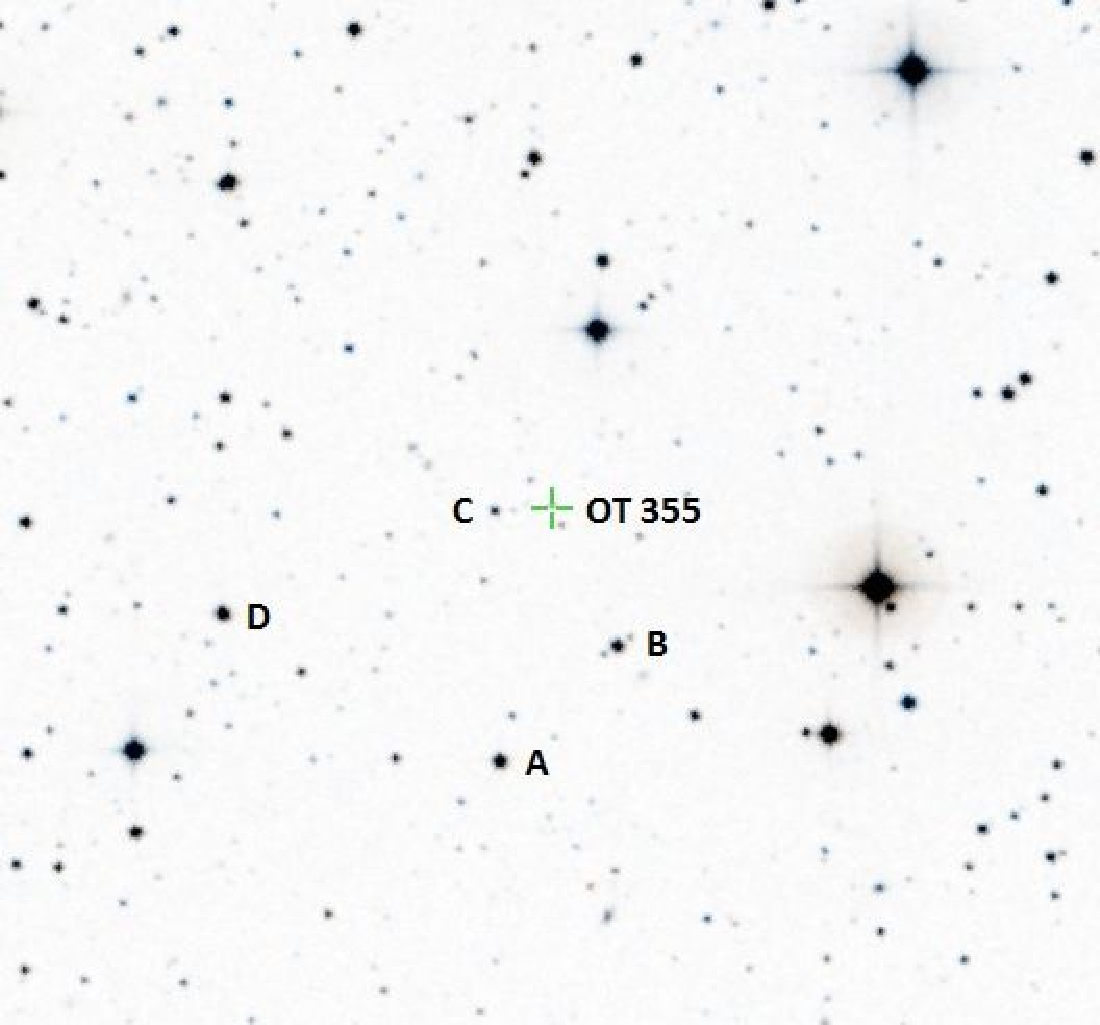}
\caption{{Secondary} 
standards in the field of OT~355, denoted as A, C, and D. {Here the object is marked with a cross and B is an additional check star}. The field is 10 arcmin wide; north is up, and east is to the left.}
\label{f5}
\end{figure}

In order to facilitate future photometric monitoring of this object, we calibrated secondary standards in the field as part of this work. The 60 cm Belogradchik telescope was used for that purpose. The standards are shown in Figure \ref{f5}, and their VRI magnitudes with the respective errors are listed in Table \ref{tab:standards}. Secondary standards in the fields of \mbox{S4~0954 + 65 \citep{Rai99}} and BL~Lac \citep{Fio96} were used for this calibration. {The transformation coefficients used for this calibration are presented in \citep{Str11}.} During several photometric nights, both calibrated and calibrating fields were observed in close time and air--mass proximity. Using the relations between the instrumental magnitudes and the already calibrated ones in each optical band, the transformation equations (which included the color dependence) were constructed. They were later used in the new field to obtain the magnitudes of the stars there. The final values were the averages from the different nights, and the uncertainties were based on the standard deviations.
\vspace{3pt}

\vspace{-9pt}
\begin{table}[H]

\caption{Magnitudes of secondary standards (Figure \ref{f5}).}
\label{tab:standards}
\begin{tabularx}{\textwidth}{C{1.7cm}C{1.5cm}C{1.5cm}C{1.5cm}C{1.5cm}C{1.5cm}C{1.5cm}}
\toprule
\textbf{Star}	&	\textbf{V}	&	\textbf{Err V}	&	\textbf{R}	&	\textbf{Err R}	&	\textbf{I}	&	\textbf{Err I}	\\
\midrule
A	&	14.39	&	0.04	&	14.04	&	0.04	&	13.73	&	0.05	\\
C	&	16.98	&	0.06	&	16.47	&	0.05	&	16.01	&	0.05	\\
D	&	13.94	&	0.04	&	13.59	&	0.04	&	13.29	&	0.05	\\
\bottomrule

\end{tabularx}
\end{table}

\subsection{Long-Term Flux and Color Variability}
\begin{figure}[H]
\includegraphics[width=100mm]{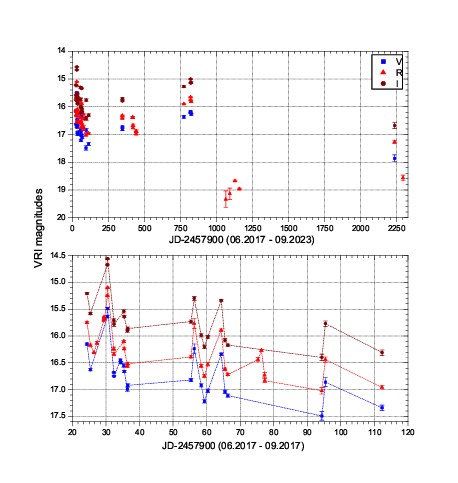}

\caption{{Long-term} 
variability in VRI colors of the blazar OT 355. The light curve covers a period of more than 3 years. The lower panel shows  the most intensive part of the monitoring in more detail, starting after the 2017 outburst. Lines are to guide the eye.}
\label{f1}
\end{figure}

\begin{figure}[H]
\includegraphics[width=70mm]{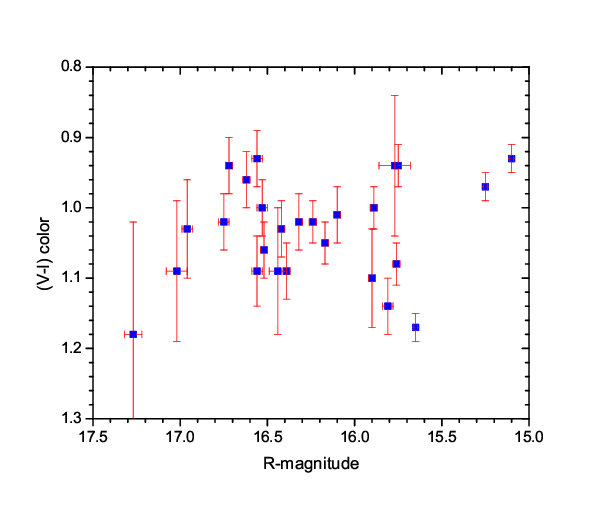}
\caption{Color--magnitude (V-I vs. R) relation for the blazar OT 355. {The seemingly apparent “bluer-when-brighter” trend is of very low statistical significance (see the text).} }
\label{f2}
\end{figure}

Figure \ref{f1} shows the long-term optical behavior of OT 355 in different colors (VRI). These magnitudes, here and below, are not corrected for the galactic absorption. Clearly, the object was very active throughout the observational period, showing variations over four magnitudes. The lower panel of Figure \ref{f1} shows  the first three months after the outburst in more detail, where one sees night-to-night changes of 0.5--0.7 mag. Figure \ref{f2} presents the color--magnitude relation for this blazar. {A very weak ``bluer-when-brighter'' trend can be found, although with a rather large scatter and often large errors. The correlation coefficient (CC) between the color and the magnitude is $0.17\pm0.16$ (a positive CC means bluer-when-brighter behavior). The errors are estimated via Monte Carlo simulations while considering Gaussian uncertainties. As a matter of fact, in $\sim$15\% of all cases, the CC was negative, and only in $\sim$10\% of the cases was the positive correlation  strong enough, so  the p-value was $<$0.05.} Note that this figure covers only the brightest states, but  during the faintest states, only the R-band magnitude was measured. It should be noted that FSRQs more often tend to show the ``redder-when-brighter'' trends at their maxima (e.g., \citep{Gu06, Gaur12}), {but no statistically significant redder-when-brighter trend was detected in this dataset.}

\vspace{-6pt}
\vspace{-9pt}

\subsection{Intra-Day Flux Variability}

\begin{figure}[H]
\includegraphics[width=130mm]{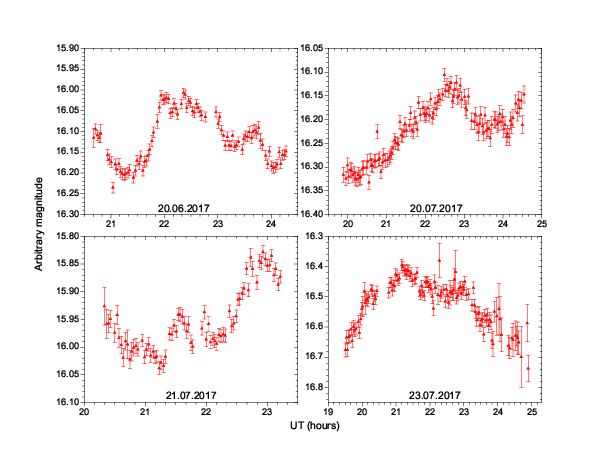}

\caption{Examples
of rapid intra-night variations in OT 355 near the outburst. Data were obtained with the 60 cm Belogradchik telescope. No filter (clear) was used to improve the signal-to-noise ratio of the light curves.}
\label{f3}
\end{figure}

The most dramatic examples of rapid intra-night variations are presented in Figure \ref{f3}, where changes of up to 0.3 mag for several hours are evident. {More variability examples are presented in the Appendix \ref{AppendixA}.}
To check any micro-variability on the intra-day timescale (IDV), we employed a statistical test, namely, the nested ANOVA test. We  employed {this} test as {it is} better than traditional F-tests and C tests \citep{Trip24}.  We  used three comparison stars, namely, A, {C}, and {D} (see Figure \ref{f5}){---a main standard (A) and two additional checks \mbox{(C and D)}.}

We also found the percentage variability amplitude. The results are presented in \mbox{Table \ref{tab:IDV}}. {Here, as we had only three comparison stars, we  used only the nested ANOVA test, as, unlike the power-enhanced F-test, it requires only $\geq$3 comparison stars \citep{Die15}.}

In the nested ANOVA test, all of the comparison stars were used as reference stars to find the differential LCs of the blazar. It partitioned the group into subgroups and then compared the means of these nested observations \citep{Die15}; then, the F-statistic was found using the ratio between the mean sum of squares (SS) within the group and the mean SS due to the nested observation of the group, as given {by} 

\begin{equation}
F = MS_{G}/MS_{O(G)}
\end{equation}
where
\begin{equation}
\begin{aligned}
MS_{G} & =\sum_{i=1}^n \frac{\left(\bar{m}_i-\bar{m}\right)^2}{a-1} \\
MS_{O(G)} & =\sum_{i=1}^n \sum_{j=1}^n \frac{\left(\bar{m}_{i j}-\bar{m}_i\right)^2}{a(b-1)}
\end{aligned}
\end{equation}

Here, $(a - 1)$ and $a(b - 1)$ are degrees of freedom ({DoF}), where $a$ is the number of groups in the night's observations, and $b$ is the number of data points in each group. At a confidence level of {99.9\%} (i.e., $\alpha$ = {0.001}), if $F > F_C$, then we say that differential LCs are variable (V); otherwise, we say that they are  non-variable (NV).

{To check for any false discoveries, we performed FDR using the Benjamini--Hochberg method \citep{Mil01}. We used the Python package (v0.14.4) {\textit{statsmodels.stats.multitest}} 
to perform FDR. In Table \ref{tab:IDV}, Yes and No indicate if the result is significant after applying FDR correction. $p_{FDR}$ indicates the \emph{p}-value after applying Benjamini--Hochberg's FDR corrections.}

\begin{table}[H]
\caption{Results of our IDV analysis of the OT 355 using a nested ANOVA test with FDR correction (\(\alpha = 0.001\)).}
\label{tab:IDV}
\begin{adjustwidth}{-\extralength}{0cm}

\centering
\begin{tabularx}{\fulllength}{C{2.2cm}C{1.2cm}C{1cm}C{1cm}C{1.5cm}C{1.5cm}C{1.5cm}C{1.8cm}C{1cm}C{1.5cm}}
\toprule
\textbf{Obs Date} & \textbf{Band} & \textbf{DoF($\nu_{1}$,$\nu_{2}$)} & \boldmath{$F$} & \boldmath{$F_c$} & \boldmath{$p$} & \boldmath{$p_\mathrm{FDR}$} & \textbf{FDR}\boldmath{$_{\alpha=0.001}$} & \textbf{Status} & \textbf{Amp (\%)} \\
\midrule
{19.06.2017}
& E & {10,33} 
& 8.48 & 4.10 & 1.2$\times10^{-6}$ 
& 2.3$\times10^{-6}$ & Yes & Var & 28.29 \\
20.06.2017 & E & 22,69 & 43.79 & 2.68 & 4.2$\times10^{-32}$ & 6.1$\times10^{-31}$ & Yes & Var & 22.19 \\
21.06.2017 & R & 13,42 & 84.96 & 3.50 & 8.1$\times10^{-26}$ & 3.4$\times10^{-25}$ & Yes & Var & 9.30 \\
22.06.2017 & E & 20,63 & 13.10 & 2.80 & 1.7$\times10^{-15}$ & 4.5$\times10^{-15}$ & Yes & Var & 19.79 \\
24.06.2017 & E & 19,60 & 47.81 & 2.87 & 3.5$\times10^{-29}$ & 2.0$\times10^{-28}$ & Yes & Var & 21.39 \\
& R & 17,54 & 63.80 & 3.03 & 1.4$\times10^{-29}$ & 1.0$\times10^{-28}$ & Yes & Var & 9.00 \\
27.06.2017 & V & 19,60 & 15.75 & 2.87 & 1.1$\times10^{-16}$ & 3.6$\times10^{-16}$ & Yes & Var & 8.90 \\
& R & 11,36 & 2.35 & 3.87 & 2.6$\times10^{-2}$ & 3.2$\times10^{-2}$ & No & NV & - \\
& I & 11,36 & 3.44 & 3.87 & 2.4$\times10^{-3}$ & 3.2$\times10^{-3}$ & No & NV & - \\
29.06.2017 & V & 6,21 & 11.18 & 5.88 & 1.3$\times10^{-5}$ & 2.1$\times10^{-5}$ & Yes & Var & 9.80 \\
20.07.2017 & E & 30,93 & 39.72 & 2.34 & 6.7$\times10^{-41}$ & 1.9$\times10^{-39}$ & Yes & Var & 22.29 \\
21.07.2017 & E & 19,60 & 52.16 & 2.87 & 3.2$\times10^{-30}$ & 3.1$\times10^{-29}$ & Yes & Var & 27.06 \\
23.07.2017 & E & 31,96 & 3.46 & 2.31 & 1.8$\times10^{-6}$ & 3.2$\times10^{-6}$ & Yes & Var & 84.76 \\
24.07.2017 & E & 5,18 & 1.13 & 6.81 & 3.8$\times10^{-1}$ & 4.1$\times10^{-1}$ & No & NV & - \\
25.07.2017 & E & 6,21 & 8.22 & 5.88 & 1.2$\times10^{-4}$ & 1.7$\times10^{-4}$ & Yes & Var & 30.49 \\
09.08.2017 & E & 35,108 & 7.09 & 2.21 & 1.6$\times10^{-15}$ & 4.5$\times10^{-15}$ & Yes & Var & 28.58 \\
10.08.2017 & E & 24,75 & 5.19 & 2.57 & 2.0$\times10^{-8}$ & 4.1$\times10^{-8}$ & Yes & Var & 19.57 \\
11.08.2017 & E & 37,114 & 14.01 & 2.17 & 6.1$\times10^{-28}$ & 2.9$\times10^{-27}$ & Yes & Var & 40.18 \\
08.05.2018 & R & 3,12 & 23.91 & 10.80 & 2.4$\times10^{-5}$ & 3.6$\times10^{-5}$ & Yes & Var & 13.56 \\
17.07.2018 & E & 20,63 & 4.36 & 2.80 & 3.8$\times10^{-6}$ & 6.4$\times10^{-6}$ & Yes & Var & 13.99 \\
19.07.2018 & E & 31,96 & 0.80 & 2.31 & 7.6$\times10^{-1}$ & 7.9$\times10^{-1}$ & No & NV & - \\
20.07.2018 & E & 14,45 & 3.75 & 3.36 & 3.7$\times10^{-4}$ & 5.1$\times10^{-4}$ & Yes & Var & 24.70 \\
10.08.2018 & E & 13,42 & 1.56 & 3.50 & 1.4$\times10^{-1}$ & 1.6$\times10^{-1}$ & No & NV & - \\
11.08.2018 & E & 12,39 & 14.38 & 3.67 & 8.6$\times10^{-11}$ & 1.9$\times10^{-10}$ & Yes & Var & 19.59 \\
08.07.2019 & R & 4,15 & 0.07 & 8.25 & 9.9$\times10^{-1}$ & 9.9$\times10^{-1}$ & No & NV & - \\
24.08.2019 & E & 13,42 & 3.02 & 3.50 & 3.3$\times10^{-3}$ & 4.2$\times10^{-3}$ & No & NV & - \\
25.08.2019 & E & 21,66 & 18.31 & 2.73 & 6.5$\times10^{-20}$ & 2.3$\times10^{-19}$ & Yes & Var & 17.50 \\
28.08.2019 & E & 21,66 & 8.78 & 2.73 & 4.4$\times10^{-12}$ & 1.1$\times10^{-11}$ & Yes & Var & 14.55 \\
28.06.2020 & R & 20,63 & 1.14 & 2.80 & 3.4$\times10^{-1}$ & 3.8$\times10^{-1}$ & No & NV & - \\
\bottomrule
\end{tabularx}
\end{adjustwidth}
\noindent{\footnotesize{Var: Variable; NV: Non-Variable. $\nu_{1}$ and $\nu_{2}$ are DoF (a-1) and a(b-1) respectively}}
\end{table}

\subsection{Intra-Day Variability Amplitude}
For each of the IDV light curves that were found to be variable (V), we computed their variability amplitude ($Amp$) using the relation given by \citep{Heidt1996}:
\begin{equation}
Amp = 100\times \sqrt{\left(A_{\max }-A_{\min }\right)^2-2 \sigma^2} \text{ (in percent)}  .
\end{equation}

{Here,} 
$A_{\max } $ \& $ A_{\min }$ are the maximum and minimum calibrated magnitudes of the blazar, respectively, and $\sigma$ is the mean error.

The shortest observed timescale is typically defined as follows:

\begin{equation}
t_{\rm var}\simeq \frac{\langle F \rangle }{|dF/dt|} \simeq \frac{1}{|dm/dt|}
\end{equation}
and can be estimated from the intra-night variations. A very sharp rise (the night of {20 June 2017,} 
Figure \ref{f3}) of about 0.2 mag for $\sim$40 min determines $t_{\rm var}\leq 2$ h ($\sim7$ ksec).

\begin{figure}[H]
\includegraphics[width=80mm]{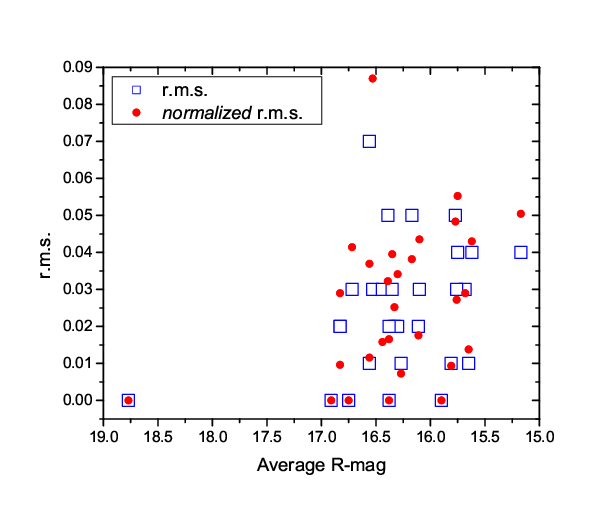}
\caption{Fractional variability--flux (``\textit{rms}--<R>'') relation for the nights for which intra-night monitoring was performed. Both \textit{rms} and time-length-normalized \textit{rms} are shown (see the text), as \textit{rms} is normally expected to increase with the length of observation. }
\label{f4}
\end{figure}

{\textls[-25]{The collection of the intra-night data allows the study of the  ``fractional variability--average flux'' relation or ``root-mean-square \endnote{Defined as 
$\sigma_{\rm rms} = \sqrt{\frac{1}{N-1}\sum_{i=1}^{N} (m_{i}-\langle m \rangle)^{2} - \langle \sigma_{\rm phot} \rangle^{2}}$,} where $m_{i}$ are the individual magnitude measurements, $\langle m \rangle$ is the average magnitude, $N$ is the number of data points, and $\langle \sigma_{\rm phot} \rangle$ is the average photometric uncertainty. $\sigma_{\rm rms}$ is taken to be zero if the expression under the square root happens to be negative}.  $\sigma_{rms}$--flux'' relation on intra-night timescales}. All data are given in Table \ref{tab:log} and presented in \mbox{Figure \ref{f1}}. The smallest $\sigma_{\rm rms}$ values should be considered as upper limits only, as the variability there may not be statistically significant (see Table \ref{tab:IDV}). \mbox{Figure \ref{f4}} shows a slight tendency of increasing $\sigma_{rms}$ with the average flux. Here, the normalized values of $\sigma_{rms}$ for the average monitoring time (2.9 h) are also shown to take into account the influence of the length of the observing series on the \textit{rms} \citep{Bach17b, Bach23}. Both quantities more or less share the same distribution in the plot. {The correlation coefficient (Pearson) between  $\sigma_{rms}$ and the average flux is similar for both definitions of $\sigma_{rms}$ and is about $-0.35$, with a \emph{p}-value of about 0.07. Spearman's rank correlation coefficients (considering one or two rather deviating points) are also similar for both quantities---$CC=-0.34\pm0.03$ with $p=0.07\pm0.03$ for $\sigma_{rms}$ and $CC=-0.39\pm0.01$ with $p=0.03\pm0.01$ for the time-normalized $\sigma_{rms}$---meaning that the correlation is significant to about a 95\% confidence level for both quantities. The $1\sigma$ errors here come from the fact that the rank of the repeating values had to be randomized.} We should point out, however, that no attempts to correct for the change in the Doppler factor of the emitting region and its influence on the time intervals have been made (e.g., \citep{Rai23}).
\subsection{Structure Function}
\begin{figure}[H]
\includegraphics[width=65mm]{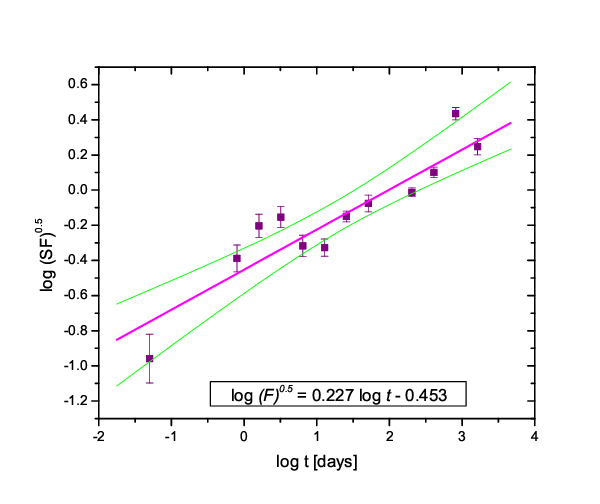}
\caption{{R-band} 
{structure} 
function of OT 355 between 2017 and 2023 ($\sim$50 data points). A linear fit with a slope of $\sim$0.23 is consistent with the data. The slope of the square root of the SF is shown as \mbox{in \citep{Kaw98}}. {The 90\% confidence intervals are also shown.}}
\label{f7}
\end{figure}

Figure \ref{f7} shows the structure function (SF) of the OT 355 R-band light curve from \citet{Sim85}; however (see also \citet{Emm10}), it can be defined \linebreak  as follows:

\begin{equation}
SF(\tau)=\frac{1}{N(\tau)}\sum_{i<j} [m(t_{\rm i})- m(t_{\rm j})]^{2},
\end{equation}
where summation is performed over all pairs in which $\tau - \Delta\tau/2 < t_{\rm j} - t_{\rm i} \leq \tau + \Delta\tau/2$, $N(\tau)$ denotes a number of such pairs, {$\tau$ is the time separation of bins, and $\Delta\tau$ is the width of each bin}. Again, no possible changes in the Doppler factor were taken into account when the time bins and/or magnitude differences were defined while building this SF (\mbox{e.g., \citep{Rai21, Rai23}).}
\vspace{-3pt}

The SF is useful for revealing how much the magnitude changes as a function of the time difference between two observations and is normally applied to unevenly sampled data. The saturation (or a slope change) of the SF might indicate the presence of some characteristic time of the variable system. Due to the combination of short- (intra-night) and long-term observations, we were able to build the SF over five orders of magnitude on the timescale. A linear slope of about {$0.23\pm0.03$} can be fitted reasonably well through the data (time is expressed in days), and no signs of saturation are evident yet for over 5 years of observations.

\subsection{Optical Spectrum}

\begin{figure}[H]
\includegraphics[width=65mm]{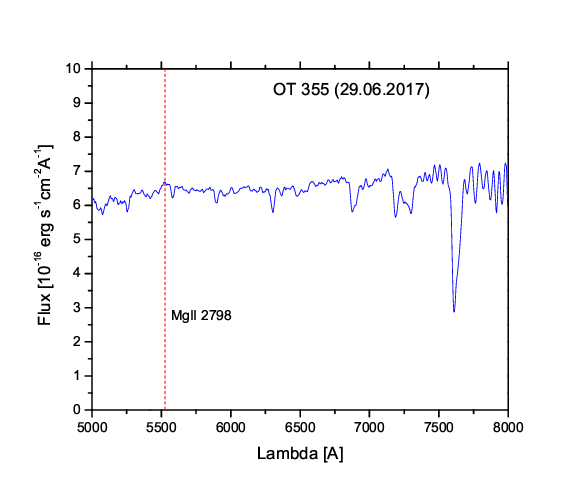}
\caption{{Optical} 
spectrum of OT 355 taken with the 2 m Rozhen telescope during the night of \linebreak  {29 June 2017}. The location of the Mg II$\lambda$2798 broad line is marked by a red dashed line.}
\label{f6}
\end{figure}

An optical spectrum covering 5000--8000 {\AA} (observer's frame) was taken during the night of 29 June 2017 with the 2 m Rozhen telescope, when OT 355 was relatively bright ($R\simeq16$ mag). {Ten} spectra with a total of about 60 min exposure time were averaged together to be shown in Figure \ref{f6}. The observing conditions were not very favorable, and the final spectrum was too noisy to be able to show fine details. The location of  Mg II $\lambda$2798 is indicated, {but, due to the insufficient S/N ratio, only an upper limit for the broad line equivalent width can be established ($EW<5$ \AA). Other researchers \citep{Xia22} reported much higher EW values (\mbox{$29\pm5$ \AA}); however, these values can be easily explained by significant continuum variations with relatively constant line luminosity.} For flux calibration, the standard BD + 33d2642 was {used}. \endnote{\url{https://www.eso.org/sci/observing/tools/standards/spectra/stanlis.html}  (accessed on 26 May 2025).} There were no de-reddening or telluric band \mbox{corrections applied.}
\vspace{-3pt}

\section{Discussion}\label{sec4}

FSRQs such as OT 355 are  known to show larger-amplitude variations {in the optical range} 
\citep{Anj20}. Unlike BL~Lacs, which show mostly bluer-when-brighter color trends,  FSRQs predominantly show the opposite trends, especially at high fluxes, due to the joint contribution of the bluer, relatively stable thermal disk emission and the redder, highly variable jet emission \citep{Neg22}. For this object, however, we found no statistically significant color--magnitude correlations (Figure \ref{f2}).

Most of the objects for which we studied the intra-night ``\textit{rms}-flux'' relation showed positive or no clear trends, e.g., S4~0954 + 65 \citep{Bach16}, CTA~102 \citep{Bach17b}, and BL~Lacertae \citep{Bach23}. One way to explain larger intra-night variations during periods of higher average fluxes could be based on the high nonlinearity of  Doppler boosting (e.g., \citep{Bach17b}). Indeed, taking into account that the emission from a blob is

\begin{equation}
I \propto D^{3+\alpha}
\end{equation}
where $D=\frac{1}{\Gamma(1-\beta \cos\theta)}$ is the Doppler factor and $\alpha\simeq1$ is the spectral index, one can assume that $\delta m \simeq \delta I/I$ would change much more significantly with a high value of $D$ (or high average fluxes). However, a more careful analysis shows that this is not exactly true. Let one consider two geometric scenarios that can account for both average flux and variability:

{(i) } 
Allow the direction of motion of a small, fixed number of emitting regions to vary rapidly with respect to the observer.

The key point is that one cannot require the direction of the jet with respect to the observer to drop to zero; rather,  the jet should pass by at a minimal angle $\theta_{\rm 0}$.  Let one assume a small angle approximation ($\theta<<1$), leading to the following:

\begin{equation}
\cos\theta \simeq 1-\frac{\theta^{2}}{2}
\end{equation}
and
\begin{equation}
D\simeq \frac{2}{\Gamma\theta^{2}}
\end{equation}
if $\beta\rightarrow1$. Then, from the relations above, one can write
\\
\begin{equation}
\delta m \simeq \frac{\delta I}{I} \propto \frac{\delta D}{D} \propto \theta \delta \theta.
\end{equation}

\begin{figure}[H]
\includegraphics[width=50mm]{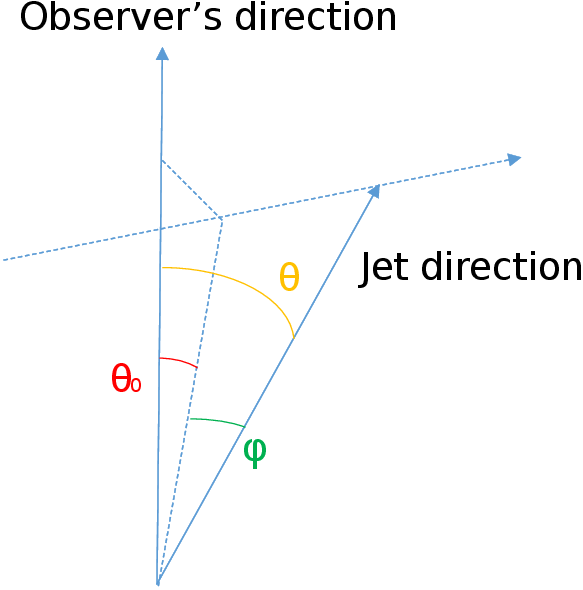}
\caption{Illustration of a random or a constant-speed change in the direction of motion of an emitting blob, passing by the direction of the observer at a closest distance of $\theta_{\rm 0}$.}
\label{f8}
\end{figure}

{Considering} that $\theta^{2} \simeq \phi^{2}+\theta_{0}^{2}$, which is approximately true for small angles ({all three angles---$\theta$, $\theta_{0}$, and $\phi$---are shown in Figure \ref{f8}}), one obtains

\begin{equation}
\theta \delta \theta \simeq \sqrt{\theta^{2}-\theta_{0}^{2}} \delta \phi
\end{equation}

\vspace{-7pt}

{Further} expressing $\theta$ with $D$ and $D$ with $I$ from the equations above, one obtains

\begin{equation}
\frac{\delta I}{I} \propto \sqrt{\left(\frac{I_{0}}{I}\right)^{1/(3+\alpha)}-1},
\end{equation}
which is a declining function of intensity. Here, the maximum intensity is $ I_{\rm 0}=I(\theta=\theta_{\rm 0})$, $I\leq I_{\rm 0}$, and $\delta \phi$ is omitted, as it can be considered a random quantity (or one that linearly changes with time), affecting $\delta I/I$ in a statistical sense only through the proportionality coefficient, which is presented on the right-hand side of the equation above. The variability is even zero at $\theta=\theta_{\rm 0}$, as there the Doppler factor is maximized, but its change at that particular point is approaching zero.

In other words, assuming a fixed number of active regions with a constantly changing direction of motion (e.g., emitting cells in a highly turbulent environment) would hardly explain the observed ``\textit{rms}-flux'' behavior for this and other similar objects unless some nonlinear interactions between the cells' activity is presumed (see, however, the TEMZ model by \citep{Mar14}).

{(ii)} Fix the direction but allow the number ($N$) of the emitting regions to vary.

If the number of the actively emitting regions can vary rapidly---for instance, due to rapid evolution---then

\begin{equation}
\frac{\delta I}{I}\simeq\frac{1}{\sqrt{N}}
\end{equation}
but $I\propto N$; therefore,
\begin{equation}
\frac{\delta I}{I}\propto\frac{1}{\sqrt{I}}
\end{equation}
which again means that the variations should decline at high fluxes.
Clearly,  \textit{{additive} 
} mechanisms---such as those briefly described above---do not provide a reliable explanation of  blazar variability, as they mostly predict the decrease in variability for high states. So, there should be some \textit{{multiplicative}
} mechanisms (e.g., avalanche-type mechanisms) involved, where a small initial disturbance can trigger a much stronger \mbox{response \citep{Asc14, Kaw98, Fan02}}. Avalanches, which are triggered by magnetic reconnections or other shock--acceleration mechanisms \citep{Kun22}, could perhaps offer an explanation for the positive (in general) \mbox{``\textit{rms}-flux'' relation.}

On the other hand, the observed significant color changes (up to 0.3 mag in the \mbox{V-I} color) suggest that, at least on longer-term timescales, a particle’s population energy evolution (due to particle acceleration and synchrotron/Compton losses)  can also play a role in the observed variability. This evolution does not necessarily drive the color variability directly, but it is enough to produce differences in the characteristics (electron densities, energy distributions, and magnetic fields) of emitting regions at different times, allowing the observed SED shape (colors) to vary in time.

The shortest observed variability timescale can be used to restrain a magnetic field if associated with the relativistic particles’ cooling time; that is,

\begin{equation}
t_{\rm cool} \leq \frac{D}{(1+z)} t_{\rm var}
\end{equation}

{Since} the synchrotron losses are due to the presence of the magnetic field, one can relate the cooling time with the magnetic field strength (e.g., \citep{Fan21}) to obtain

\begin{equation}
B \leq 3 (t_{\mathrm var}, [h])^{-2/3} \mathrm ~\rm{G},
\end{equation}
which is approximately accurate up to a small coefficient and takes  the additional Compton losses into account. For the case of OT 355, we get $B \simeq 2 \mathrm G$. In addition, the shortest timescale restricts the size of the emission region as follows:

\begin{equation}
R\leq\frac{D}{1+z}ct_{\rm var},
\end{equation}
which can be roughly estimated to be $R$$\sim$$10^{15}$ cm, provided $D$$\sim$$10$. This estimate can be of use for further modeling of the SED.

Strong emission lines are not expected to dominate the blazar optical spectrum (\mbox{Figure \ref{f7}}), even though OT 355 is classified as an FSRQ. At maximum light, the non-thermal component significantly dominates the spectrum, and the lines are likely concealed in the much stronger continuum and the noise. In addition, further suppression of the broad emission can occur for strong jets that are pointed directly towards the observer. Then, the entire broad-line region would be visible entirely through the jet cone, and a significant part of its emission could be lost due to Compton scattering \citep{Bach17b}.

Somewhat surprising is the lack of saturation of the SF for long, as most  blazars’ SFs show a plateau after a year or so (e.g., \citep{Zha12}). Apparently, this particular object displays a more complex jet evolution, which is perhaps linked to slower processes, such as the change in the accretion rate or the jet precession, both of which may be caused by the presence of a BH binary (we have no evidence for this, however). The SF also appears to be flatter when compared with the slope of other blazars. The SF slope, which is defined as

\begin{equation}
\beta = \frac{ d \log \rm \sqrt{SF}}{ d\log t}
\end{equation}
varies between 0 (flicker noise) and 1 (shot noise) for most accreting astrophysical \mbox{objects \citep{Kaw98}}. For instance, \citet{Rai21} found $\beta\simeq1$ for S5~0716 + 714 based on TESS data of a length of 21 days, and \citet{Zha13} reported a slope of 0.63 (R-band) for the long-term light curve of PKS~0537-441. On much shorter scales ($<$3 h), \citet{Bach17b} found a slope of $\simeq$0.4 for CTA~102.

\section{Conclusions}
We presented the results of long-term and intra-night optical studies of the highly variable blazar OT 355. For this object, we found an unusually flat structure function for the long-term variations with no apparent saturation. No conclusive color--magnitude or rms--flux relations were found. We show that a negative \textit{rms}--flux tendency should have been observed if additive geometric reasons drove the short-term variations. Using the shortest variability timescale observed, we were able to restrain the magnetic field of the jet and the typical size of an emitting region.

Blazar monitoring, especially at different timescales and in different colors, can provide unique insights into the physics of relativistic jets. Optical data, even from smaller telescopes, can be an important addition to the high-energy data from ground-based and space instruments, and they can help to better understand the physics of jetted AGNs.

\vspace{6pt}
\authorcontributions{R.B. wrote most of the text and performed many of the observations and reductions. T.T. and A.C.G. added to the text and performed the statistical tests. A.K., Y.N., S.B., G.L., B.S., M.M. and E.O. contributed to the observations. A.S., W.-X.Y., Y.L. and J.-H.F. suggested many changes to improve the text. All authors have read and agreed to the published version of \mbox{the manuscript.}}

\funding{{This} 
research was partially supported by the Bulgarian National Science Fund of the Ministry of Education and Science under grants KP-06-H68/4 (2022), KP-06-H78/5 (2023), KP-06-KITAJ/12 (2024), and KP-06-H88/4 (2024). J.H. Fan's work was partially supported by the National Natural Science Foundation of China (NSFC 12433004) and the Eighteenth Regular Meeting Exchange Project of the Scientific and Technological Cooperation Committee between the People's Republic of China and the Republic of Bulgaria (Series No. 1802).}




\dataavailability{All published data are available upon request.}

\conflictsofinterest{The authors declare no conflicts of interest. The funders had no role in the design of the study, in the collection, analyses, or interpretation of data, in the writing of the manuscript, or in the decision to publish the results.}

\appendixtitles{yes}
\appendixstart
\appendix

\begingroup\makeatletter\def\f@size{12}\check@mathfonts
\def\maketag@@@#1{\hbox{\m@th\fontsize{10}{10}\selectfont\normalfont#1}}

\section[\appendixname~\thesection]{{Intra-Night} 
Variability Examples}\label{AppendixA}
\endgroup

\vspace{-6pt}
\begin{figure}[H]
\includegraphics[width=115mm]{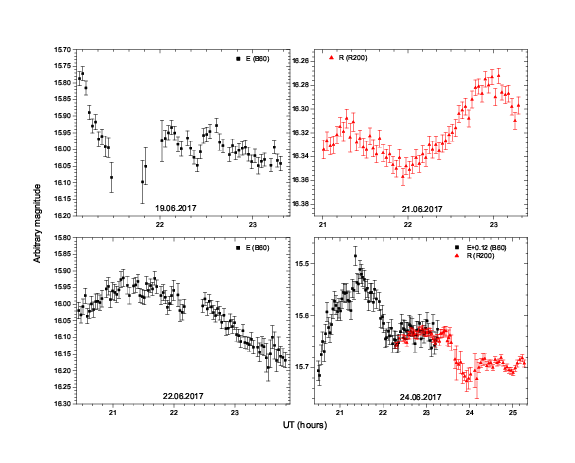}
\caption{{Examples}
of the intra-night variability of OT~355. The band in use, as well as the telescopes, are indicated for each night (see the main text). Filter E indicates unfiltered light.}
\label{a1}
\end{figure}
\vspace{-6pt}
\begin{figure}[H]
\includegraphics[width=115mm]{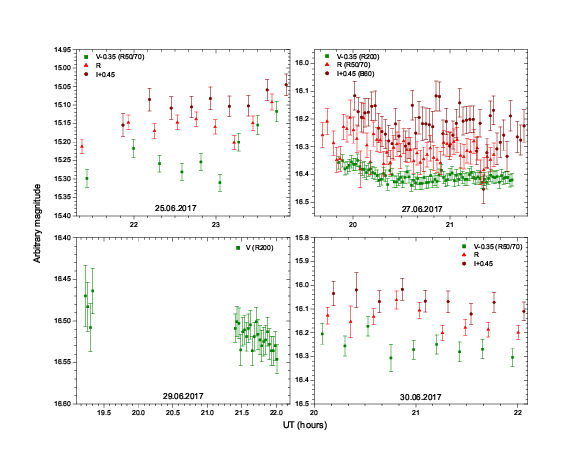}
\caption{{See} 
Figure \ref{a1}.}
\label{a2}
\end{figure}
\vspace{-6pt}
\begin{figure}[H]
\includegraphics[width=115mm]{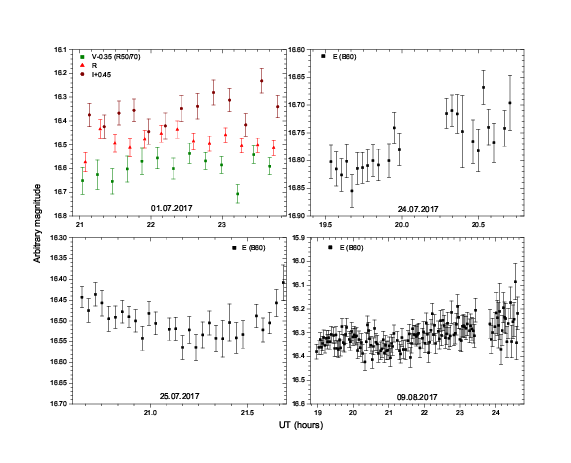}
\caption{{See} Figure \ref{a1}.}
\label{a3}
\end{figure}
\vspace{-6pt}
\begin{figure}[H]
\includegraphics[width=115mm]{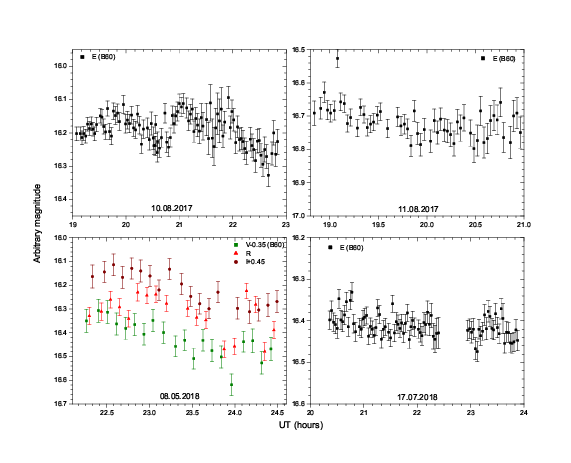}
\caption{{See} Figure \ref{a1}.}
\label{a4}
\end{figure}
\vspace{-6pt}
\begin{figure}[H]
\includegraphics[width=115mm]{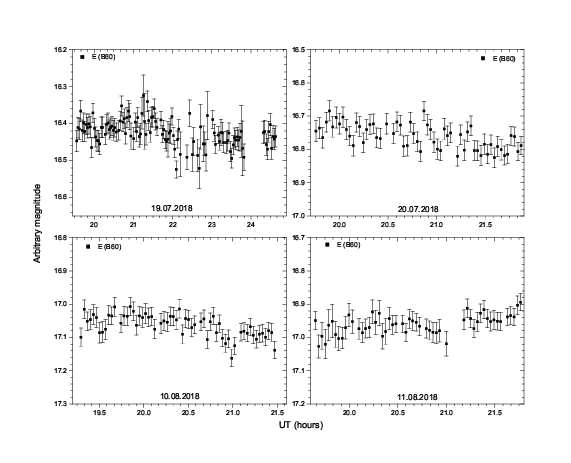}
\caption{{See} Figure \ref{a1}.}
\label{a5}
\end{figure}
\vspace{-6pt}
\begin{figure}[H]
\includegraphics[width=115mm]{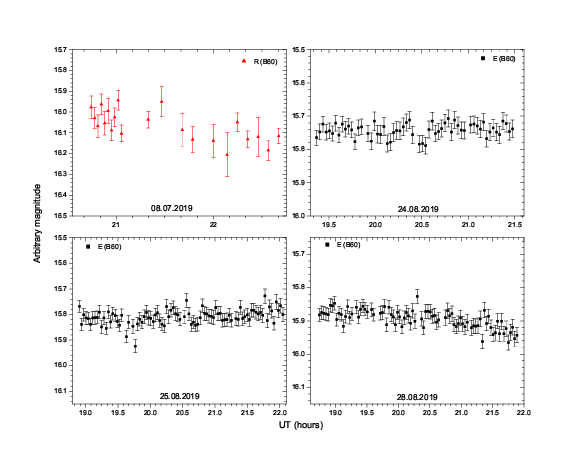}
\caption{{See} Figure \ref{a1}.}
\label{a6}
\end{figure}
\vspace{-6pt}
\begin{figure}[H]
\includegraphics[width=90mm]{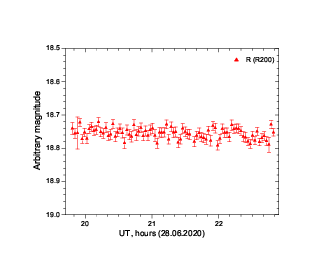}
\caption{{See} Figure \ref{a1}.}
\label{a7}
\end{figure}

\printendnotes

\reftitle{References}
\begin{adjustwidth}{-\extralength}{0cm}

%
\PublishersNote{}
\end{adjustwidth}

\end{document}